\documentclass[aps,prd,amsfonts,nofootinbib,twocolumn]{revtex4} 
\usepackage{graphics,epsfig}

\begin{document}
\preprint{WM-03-113}
\title{A Naturally Narrow Positive Parity $\Theta^+$}
\author{Carl E. Carlson} \email[]{carlson@physics.wm.edu}
\author{Christopher D. Carone}\email[]{carone@physics.wm.edu}
\author{Herry J. Kwee}\email[]{herry@camelot.physics.wm.edu}
\author{Vahagn Nazaryan}\email[]{vrnaza@wm.edu}
\affiliation{Particle Theory Group, Department of Physics,
College of William and Mary, Williamsburg, VA 23187-8795}
\date{December 2003}
\begin{abstract}
We present a consistent color-flavor-spin-orbital wave function for a
positive parity $\Theta^+$ that naturally explains the observed
narrowness of the state.  The wave function is totally symmetric in
its flavor-spin part and totally antisymmetric in its color-orbital
part.  If flavor-spin interactions dominate, this wave function
renders the positive parity $\Theta^+$ lighter than its negative
parity counterpart.  We consider decays of the $\Theta^+$ and compute
the overlap of this state with the kinematically allowed final states.
Our results are numerically small.  We note that dynamical
correlations between quarks are not necessary to obtain narrow
pentaquark widths.  
\end{abstract}
\maketitle

%\vglue -7.3cm \noindent WM-03-113; hep-ph/0312325  \vglue 6.4cm

%%%%%%%%%%%%%%%%%%%%%%%%%%%%%%%%%%%%%%%%%%%%%%%%%%%%%%%%%%%%%%%%%%%%
%\section{Introduction}\label{sec:intro}
%%%%%%%%%%%%%%%%%%%%%%%%%%%%%%%%%%%%%%%%%%%%%%%%%%%%%%%%%%%%%%%%%%%%

{\em Introduction.} The recent discovery of pentaquark
states~\cite{nakano,barmin,stepanyan,barth,kubarovsky,asratyan,Alt:2003vb}
has stimulated a significant body of theoretical~[8--37] and
experimental research.  Pentaquarks are baryons whose minimal Fock
components consist of four quarks and an antiquark.  The first
observed pentaquark was the $\Theta^+(1540)$ with strangeness $S=+1$,
and with quark content $udud\bar s$. More recently, the NA49
Collaboration~\cite{Alt:2003vb} has reported a narrow
$\Xi^{--}_5(1860)$ baryon with $S=-2$ and quark content $dsds\overline
u$, together with evidence for its isoquartet partner $\Xi^{0}_5$ at
the same mass.

The existence of the $\Theta^+$, as well as its flavor quantum
numbers, seems to be well established (for a different view, see
Ref.~\cite{sze}).  If the $\Theta^+$ were a member of an isovector or
isotensor multiplet, then one would expect to observe its doubly
charged partner experimentally. The SAPHIR Collaboration~\cite{barth}
searched for a $\Theta^{++}$ in $\gamma p \rightarrow \Theta^{++} K^-
\rightarrow pK^+K^-$, with negative results.  They concluded that the
$\Theta^+$ is an isosinglet and hence a member of a pentaquark
antidecuplet.  All but one theoretical paper~\cite{capstick} treat the
$\Theta^+$ as an isosinglet.

The spin and parity quantum numbers of the $\Theta^+$ have yet to be
determined experimentally.  The spin of $\Theta^+$ is taken to be 1/2
by all theory papers to our knowledge and various estimates show that
spin-3/2 pentaquarks must be heavier~\cite{csm}.  A more controversial
point among theorists is the parity of the state.  For example, QCD
sum rule calculations~\cite{QCD_sum_rule}, quenched lattice
QCD~\cite{lattice_QCD}, and a minimal constituent quark treatment by
the present authors~\cite{us}, predict that the lightest $\Theta^+$ is
a negative parity isosinglet. All chiral soliton
papers~\cite{diakonov, weigel}, some correlated quark
models~\cite{jaffew,karlinerl}, and some works within the constituent
quark model~\cite{stancu,glozman,jennings} including a second work by
the present authors~\cite{ppppp}, predict the lightest $\Theta^+$
pentaquark as a positive parity isosinglet.

The photoproduction and the pion-induced production cross sections of
the $\Theta^+$ were studied in~\cite{oh_pro}.  It was shown in both
cases that the production cross sections for a negative parity
$\Theta^+$ are much smaller than those for the positive parity state
(for a given $\Theta^+$ width).  In Ref.~\cite{oh_pro}, results for
the $\Theta^+$ production cross section in photon-proton reactions
were compared with estimates of the cross section based on data
obtained by the SAPHIR Collaboration~\cite{barth}, and odd-parity
pentaquark states were argued to be disfavored.

Here, we present new results following from a consistent treatment of
the color-flavor-spin-orbital wave function for a positive parity
$\Theta^+$.  In~\cite{ppppp} (inspired by \cite{stancu}), we showed
that dominant flavor-spin interactions render the positive parity
$\Theta^+$ lighter than its negative parity counterpart.  Here we will
present decompositions of the quark model wave function of the
$\Theta^+$, explicitly including the orbital part.  We will see that
the narrowness of the $\Theta^+$ follows naturally from the group
theoretic structure of the state.

%%%%%%%%%%%%%%%%%%%%%%%%%%%%%%%%%%%%%%%%%%%%%%%%%%%%%%%%%%%%%%%%%%%%
%\section{Wave Function}\label{sec:wf}
%%%%%%%%%%%%%%%%%%%%%%%%%%%%%%%%%%%%%%%%%%%%%%%%%%%%%%%%%%%%%%%%%%%%

{\em Wave Function.} If flavor-spin interactions dominate~\cite{gr},
the lightest positive parity $\Theta^+$ will have a flavor-spin (FS)
wave function that is totally symmetric~\cite{stancu,jennings,ppppp}.
Fermi-Dirac statistics dictates that the color-orbital (CO) wave
function must be fully antisymmetric. We present two decompositions of
the wave function, one in terms of quark pairs and the antiquark, and
another in terms of the quantum numbers of $q^3$ and $q \bar q$
subsystems.

In the first decomposition, the overall $q^4$ flavor state must be a
${\bf\bar 6}$.  This is the only representation that one can combine
with a flavor ${\bf \bar 3}$ (the antiquark) to form an antidecuplet.
This further implies that the overall $q^4$ spin is 0, since the only
possible fully symmetric $q^4$ $(F,S)$ wave functions are $({\bf \bar
6},0)$ or $({\bf 15}_{\rm M}, 1)$.  A flavor ${\bf \bar 6}$ can be
obtained if both quarks pairs are in either a {\bf 6} or ${\bf \bar
3}$, while a spin-0 state can be obtained if both are either spin-0 or
1.  Since we want a fully symmetric FS wave function, we must combine
these possibilities as follows:
\begin{equation}
\left| FS \right\rangle_{(\bar{\mathbf{6}},0)} = a \left\vert (\bar
{\bf 3},0)(\bar {\bf 3},0) \right\rangle_{(\bar{\bf 6}, 0)} + b
\left\vert ({\bf 6},1)({\bf 6},1) \right\rangle_{(\bar{\bf 6},0)} \ .
\label{FS1}
\end{equation} 
The parentheses on the right hand side delimit the flavor and spin
quantum numbers of the first and second pair of quarks, each of which
is combined into an overall $(\bar{\bf 6},0)$.  For the $\Theta^+$,
the $q^4$ states on the right-hand-side are:
\begin{eqnarray}
\left\vert (\bar {\bf 3},0)(\bar {\bf 3},0) \right \rangle_{(\bar{\bf
			6},0)} &=& \nonumber \\ \frac{1}{4}(ud-du)(ud-
			&& {\kern -1.8em} du) \otimes
			(\uparrow\downarrow-
			\downarrow\uparrow)(\uparrow\downarrow-\downarrow\uparrow)
			\nonumber \ , \\[1.3ex] \left\vert ({\bf
			6},1)({\bf 6},1) \right\rangle_{(\bar{\bf
			6},0)} &=& \\ \nonumber \frac{1}{12}
			(2uudd+2dduu&-&udud-uddu - duud-dudu) \\
			\nonumber \otimes \big(
			2\uparrow\uparrow\downarrow\downarrow
			+2\downarrow\downarrow\uparrow\uparrow &-&
			\uparrow\downarrow\uparrow\downarrow-
			\uparrow\downarrow\downarrow\uparrow
			-\downarrow\uparrow\uparrow\downarrow-
			\downarrow\uparrow\downarrow\uparrow \big) \ .
\label{FS2}
\end{eqnarray}

\noindent Total symmetry of the wave function demands $a$ = $b$. To
properly normalize the state, we choose $a=b= 1/{\sqrt 2}$.

\begin{figure}[t]
\epsfxsize 3.3 in \epsfbox{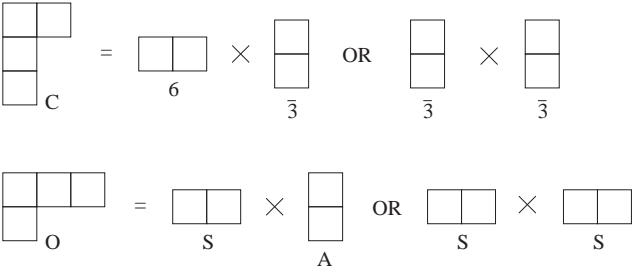} \caption{All possible states
that can be appropriately combined to yield a totally antisymmetric CO
state}
\label{fig:tableaux}
\end{figure}

The next step is to construct the totally antisymmetric CO wave
function.  The $q^4$ color state must be a ${\mathbf 3}$, which is a
mixed symmetry state, whose Young tableaux is shown in
Fig.~\ref{fig:tableaux}.  The orbital state, containing three
$S$-states and one $P$-state, must have a permutation symmetry given
by the conjugate tableaux in order to obtain overall
antisymmetry. Hence the structure of our wave functions implies that
the strange antiquark is not orbitally excited; simple estimates
suggest that a state with the $\bar s$ excited would be considerably
heavier~\cite{ppppp}.  The possible color and orbital representations
for two pairs of quarks are shown in Fig.~\ref{fig:tableaux}.

%%%%%%%%%%%%%%%%%%%%%%%%%%%%%%%%%%

%%%%%%%%%%%%%%%%%%%%%%%%%%%%%%%%%%%

From Fig.~\ref{fig:tableaux}, a totally antisymmetric CO wave function
must have the form:
\begin{eqnarray}
\left| CO \right\rangle &=& a' \left\vert (\bar {\bf 3},\bf{S})(\bar
{\bf 3},\bf{S}) \right\rangle \nonumber \\ &+& b' \{\left\vert ({\bf
6},\bf{A})(\bar{\bf 3},\bf {S}) \right\rangle + \left\vert (\bar{\bf
3},\bf {S})({\bf 6},\bf{A}) \right\rangle\}\ .
\label{CO1}
\end{eqnarray}

\noindent The coefficients $a'$ and $b'$ are fixed by the constraint
that the wave function must be antisymmetric under interchange of the
first and third quarks.  When the $q^4$ color state is red, the
explicit expressions for the wave functions on the right-hand-side
are:
\begin{eqnarray}
\left\vert (\bar {\bf 3},\bf{S})(\bar {\bf 3},\bf{S}) \right\rangle
&=& \frac{1}{\sqrt 8} \big\{(RG-GR)(BR-RB) \nonumber \\ && \qquad -
(BR-RB)(RG-GR) \big\} \nonumber \\ &\otimes& \frac{1}{2}
\big\{SS(SP+PS)-(SP+PS)SS \big\} \ , \nonumber \\ \left\vert ({\bf
6},\bf{A})(\bar{\bf 3},\bf {S}) \right\rangle &=& \frac{1}{4}
\big\{(2RR(GB-BG) \nonumber \\ +(RG+GR&& {\kern -1.9em}) (BR - RB) +
(RB+BR)(RG-GR) \big\} \nonumber \\ &\otimes& \frac{1}{\sqrt
2}(SP-PS)SS \ .
\label{CO2}
\end{eqnarray}

\noindent The wave function is properly normalized with the choice
$a'=b'= 1/{\sqrt 3}$.  In our construction, the total spin of the $q^4
\bar q$ can only be $1/2$.  Appropriate Clebsch-Gordan coefficients
may be chosen to combine the orbital angular momentum of the excited
$q$ so that the total $\Theta^+$ spin is $1/2$.  We leave this
implicit.

For the second decomposition, we note that the $q^3$ and $q \bar q$
flavor wave functions must both be $\mathbf 8$'s if one is to form a
flavor $\overline{\bf 10}$.  Since the $q^4$ FS wave function is fully
symmetric, the $q^3$ FS wave function must be fully symmetric also.
The mixed symmetry of the $q^3$ flavor wave function implies that the
$q^3$ spin wave function must have mixed symmetry also and hence is
spin-1/2.  Total symmetrization of the $q^3$ FS wave function is
obtained as follows:
\begin{equation}
\left| \left({\mathbf 8}, 1/2 \right) \right\rangle_{q^3} =
		\frac{1}{\sqrt 2}\Big[ \left| \left( {\bf 8}_S, 1/2_S
		\right) \right\rangle + \left| \left( {\bf 8}_A, 1/2_A
		\right) \right\rangle \Big] \ ,
\end{equation}

\noindent where %the subscripts $S$ and $A$ refer to the permutation
symmetry of the first two quarks.  The $q\bar q$ spin can be 0 or 1.
The fully symmetric FS wave function is of the form
\begin{eqnarray}
\left| FS \right\rangle_{(\overline{\mathbf{10}},1/2)} &=& a''
\left\vert ({\bf 8},1/2)({\bf 8},0)
\right\rangle_{(\overline{\mathbf{10}},1/2)} \nonumber \\ &+& b''
\left\vert ({\bf 8},1/2)({\bf 8},1)
\right\rangle_{(\overline{\mathbf{10}},1/2)} \ ,
\label{FS3}
\end{eqnarray}     

\noindent where the coefficients $a''$ and $b''$ are fixed by
requiring that the wave function is symmetric under the interchange of
the first and fourth quarks.  For the $\Theta^+$, the part of the
states on the right-hand-side that have $z$-component spin projection
$1/2$ are:
\begin{eqnarray}
\left\vert ({\bf 8},1/2)({\bf 8},0)
	\right\rangle_{(\overline{\mathbf{10}},1/2)} &=& \nonumber
	\\[1.2ex] \frac{1}{\sqrt 2}
	\bigg[\frac{1}{4}(ud-du)(ud-du)\bar s &\otimes&
	(\uparrow\downarrow-\downarrow\uparrow)\uparrow
	(\uparrow\downarrow-\downarrow\uparrow) \nonumber \\[1.2ex] +
	\frac{1}{12}(2uudd+2dduu - udud &-& uddu-duud -dudu)\bar s
	\nonumber \\ \otimes (2\uparrow\uparrow\downarrow -
	\uparrow\downarrow\uparrow &-&\downarrow\uparrow\uparrow)
	(\uparrow\downarrow-\downarrow\uparrow) \bigg] \ ,
\label{FS4}
\end{eqnarray} 

\noindent and
\begin{eqnarray}
\left\vert ({\bf 8},1/2)({\bf 8},1)
	\right\rangle_{(\overline{\mathbf{10}},1/2)} &=&
	\frac{1}{\sqrt 2} \bigg[\frac{1}{2}(ud-du)(ud-du)\bar s
	\nonumber \\ \otimes \ \Big\{ \frac{1}{\sqrt 3}
	(\uparrow\downarrow-\downarrow\uparrow)
	\downarrow\uparrow\uparrow &-& \frac{1}{\sqrt {12}}
	(\uparrow\downarrow-\downarrow\uparrow)
	\uparrow(\uparrow\downarrow+\downarrow\uparrow) \Big\}
	\nonumber \\ + \frac{1}{\sqrt {12}} (2uudd+2dduu &-&
	udud-uddu-duud \nonumber \\ -dudu)\bar s \ \otimes \ \Big\{
	\frac{1}{3} ( && {\kern -1.9em} \uparrow\downarrow\downarrow +
	\downarrow\uparrow\downarrow-2\downarrow\downarrow\uparrow)\uparrow\uparrow
	\nonumber \\ -\frac{1}{6} (2\uparrow\uparrow\downarrow -
	\uparrow\downarrow\uparrow &-& \downarrow\uparrow\uparrow)
	(\uparrow\downarrow+\downarrow\uparrow) \Big\} \bigg] \ .
\label{FS5}
\end{eqnarray} 

\noindent These are sufficient to show $a''=-1/2$ and $b''=-\sqrt
3/2$, using a sign convention consistent with our previous
decomposition.

The CO wave function includes two possibilities.  Either the orbital
wave function is totally symmetric, $|CO\rangle_1$, or it has mixed
symmetry, $|CO\rangle_2$, and the full wave function is
\begin{equation}
|CO\rangle = |CO\rangle_1 + |CO\rangle_2 \ .
\label{CO}
\end{equation}

\noindent For the totally symmetric orbital part, one has
\begin{eqnarray}
|CO\rangle_1 &=& \frac{1}{\sqrt {18}}\{\epsilon_{ijk}C^iC^jC^k\}
	\{C^l\bar C_l\} \otimes \\ \nonumber \Big\{ a''' (SSSPS) &+&
	b''' \frac{1}{\sqrt 3}(SSP+SPS+PSS)SS \Big\} \ ,
\end{eqnarray}

\noindent where we note that the $P$-state quark can be in either the
$q^3$ or the $q \bar q$ part, and that the color wave function for the
$q^3$ part is totally antisymmetric.  The second possibility is that
the $q^3$ orbital wave function has mixed symmetry and includes the
$P$-state quark.  The mixed symmetry orbital wave function may be
either symmetric (${\mathbf M}_{\rm S}$) or antisymmetric (${\mathbf
M}_{\rm A}$) under interchange of the first two quarks.  These states
combine with color ${\bf 8}_S$ or ${\bf 8}_A$ states as $[ ({\mathbf
M}_{\rm S}, {\bf 8}_A ) - ({\mathbf M}_{\rm A}, {\bf 8}_S )
]/\sqrt{2}$, to have a fully antisymmetric $q^3$ CO wave function.  In
this case the $q\bar q$ must be a color octet and its orbital part is
symmetric.  Thus,
\begin{eqnarray}
|CO\rangle_2 &=& \frac{c'''}{\sqrt 2}\bigg[\frac{1}{\sqrt2}(SP-PS)SSS
  \nonumber\\ &\otimes& \frac{1}{4\sqrt 3} \Big\{(C^iR+RC^i)(GB-BG)
  \nonumber\\ && \qquad + \ (C^iG+GC^i)(BR-RB) \nonumber\\ && \qquad +
  \ (C^iB+BC^i)(RG-GR) \Big\}\bar C_i \nonumber\\ &+& \Big\{
  \frac{1}{\sqrt 6}(SP+PS)SSS-\sqrt{\frac{2}{3}}SSPSS \Big\}
  \nonumber\\ &\otimes& \frac{1}{12} \Big\{2(GB-BG)C^iR \nonumber\\
  &&\quad + \ 2(BR-RB)C^iG+2(RG-GR)C^iB \nonumber\\ &&\quad + \
  \epsilon^{ijk}\epsilon_{jlm} \epsilon_{krs}C^lC^mC^rC^s \Big\} \bar
  C_i \bigg] \ .
\end{eqnarray}

\noindent The above wave function is antisymmetric by construction
under the interchange of the first three quarks.  The coefficients
$a'''$, $b'''$, and $c'''$ are found to be $1/2$, $-1/\sqrt {12}$, and
$\sqrt{2/3}$, respectively, by antisymmetrizing on the first and
fourth quarks.
%It can be shown after some reorganization that the wave
%function $|CO\rangle$ is consistent with our previous CO wave function,
%Eq.~(\ref{CO1}).

%%%%%%%%%%%%%%%%%%%%%%%%%%%%%%%%%%%%%%%%%%%%%%%%%%%%%%%%%%%%%%%%%%%%

%\section{Narrow Width}\label{sec:nw}

%%%%%%%%%%%%%%%%%%%%%%%%%%%%%%%%%%%%%%%%%%%%%%%%%%%%%%%%%%%%%%%%%%%%

{\em Narrow Width.} A narrow $\Theta^+$ width can be understood if the
overlap of the color-flavor-spin-orbital wave function
%, the product of Eqs.~(\ref{FS3}) and~(\ref{CO}), 
with an $NK$ final state is numerically small.  The relevant piece of
the FS wave function is $\left\vert ({\bf 8},1/2)({\bf 8},0)
\right\rangle$, which has coefficient $a''=-1/2$.  The relevant piece of
the CO wave function has both the $q^3$ and $q \bar q$ in their
relative ground states and has each of them separately color singlet.
Furthermore, the terms of interest in the orbital wave function are
totally symmetric in their $q^3$ and $q \bar q$ parts separately.
These terms may be read from,
\begin{eqnarray}
&& \frac{a'''}{\sqrt 2} (SSS) \left\{ \frac{1}{\sqrt
2}(PS+SP)+\frac{1}{\sqrt 2}(PS-SP) \right\} \nonumber \\ &+& b'''
\frac{1}{\sqrt 3}(SSP+SPS+PSS)SS \ .
\end{eqnarray}

\noindent The totally symmetric orbital wave functions with a
$P$-state included correspond to a ground state baryon or meson with
center-of-mass motion.  From the previous section we know $a''' = 1/2$
and $b''' = -1/\sqrt{12}$.  Hence the total probability of the
$\Theta^+$ overlap with $NK$ is:
\begin{equation}
c_+ =
\left(a'' a'''/\sqrt{2}\right)^2 +
\left(a'' b'''\right)^2 =
\frac{5}{96} \ ,
\end{equation}

\noindent which implicitly includes a sum over $z$-component spin
projections.  This is interestingly small. The $\Theta^+$ width
for a positive ($\Gamma_+$) or negative ($\Gamma_-$) parity state
is
\begin{eqnarray}
\Gamma_\pm &=& c_\pm\, g_\pm^2\cdot \frac{M}{16\pi} 
\left[(1-\frac{(m+\mu)^2}{M^2})
(1-\frac{(m-\mu)^2}{M^2})\right]^{1/2} \nonumber \\
&\times& \left[(1\mp\frac{m}{M})^2-\frac{\mu^2}{M^2}\right] \,\,\, ,
\end{eqnarray}
where $M$, $m$ and $\mu$ are the masses of the $\Theta^+$, the final
state baryon and the meson, respectively, $c_\pm$ is the dimensionless
spin-flavor-color-orbital overlap factor ($c_+=5/96$, or
$c_-=1/4$ from Ref.~\cite{us}), and $g_\pm$ is an effective meson-baryon
coupling constant, ${\cal L}_{eff}(c_\pm=1)=g_{-} \bar N K^\dagger \Theta^+$ or $i g_+
\bar N \gamma^5 K^\dagger \Theta^+$.  Applying the rules of naive
dimensional analysis (NDA)~\cite{nda}, one estimates that $g_\pm \sim
4 \pi$, up to order one factors.  One then finds
\begin{equation}
\Gamma_+ \approx 4.4\mbox{ MeV} 
\mbox{  while  }
\Gamma_- \approx 1.1\mbox{ GeV}.
\end{equation}
In the effective theory approach, effects associated with long-distance
dynamics are subsumed in the values of the couplings $g_\pm$.  For example,
an explicit computation of quark wave function overlaps in baryons with both 
S- and P-wave constituents could lead to a smaller estimate for $g_+$. 
However, the precise outcome is strongly model dependent and we do not  
pursue this issue further.  Our result implies that a positive parity 
$\Theta^+$ is narrow, independent of these uncertainties.

It has been noted~\cite{jennings} that the correlated diquark state
advocated in Ref.~\cite{jaffew} has a small overlap with the $NK$
state, even if one just considers the color-flavor-spin wave function.
However, the $q^4$ part of the correlated diquark state presented
in~\cite{jaffew} is not perfectly antisymmetric.  The state is a good
approximation to a Fermi-Dirac allowed state only to the extent that
the diquarks are significantly more compact than the overall
state. The significant likelihood that the diquarks are comparable in
size to the entire pentaquark is reason for concentrating on a
consistent, antisymmetrized wave function. (We can nonetheless report
for the correlated diquark model that inclusion of the orbital wave
function reduces the $\Theta^+$ overlap with $NK$ from the
Jennings-Maltman~\cite{jennings} color-flavor-spin result of $1/24$ to
a remarkably small $5/576$.)

%%%%%%%%%%%%%%%%%%%%%%%%%%%%%%%%%%%%%%%%%%%%%%%%%%%%%%%%%%%%%%%%%%%%
%\section{Conclusions}\label{sec:conc}
%%%%%%%%%%%%%%%%%%%%%%%%%%%%%%%%%%%%%%%%%%%%%%%%%%%%%%%%%%%%%%%%%%%%

{\em Conclusions.} We have presented an explicit framework in which
the width of a positive parity $\Theta^+$ is narrow.  We find that the
spin-flavor-color-orbital overlap probability for decays to kinematically 
allowed final states is $5/96$. By comparison, the same overlap probability 
for the negative parity case is $1/4$, as was shown in Ref.~\cite{us}. Without 
any incalculable dynamical suppression (that could render $g_-$ substantially 
less than $g_+$ above), one may infer that a negative 
parity pentaquark state, if it exists, is significantly broader than its 
positive parity cousin.  Aside from its $NK$ component, the even parity 
$\Theta^+$ wave function overlaps with other color-singlet-color-singlet 
baryon-meson states that are together heavier than the $\Theta^+$, and
with color-octet-color-octet baryon-meson states.  Hence, even though the 
decay proceeds via a fall-apart mode, the amplitude to kinematically allowed 
baryon-meson states is small.
%%%%%%%%%%%%%%%%%%%%%%%%%%%%%%%%%%

\begin{acknowledgments}
We thank Frank Close for remarks and the NSF for support under Grant
Nos.\ PHY-0140012, PHY-0243768, PHY-0352413, and PHY-0245056.  CDC
thanks the William and Mary Endowment Association for its support.
HJK thanks the DOE for support under contract DE-AC05-84ER40150.
\end{acknowledgments}

%%%%%%%%%%%%%%%%%%%%%%%%%%%%%%%%%%%

\end{document}